\documentclass[nofootinbib,amsmath,amssymb,amsfonts,aps,prd,eqsecnum,superscriptaddress]{revtex4-1}

\pdfoutput=1
\usepackage{hyperref}
\usepackage{amsmath}
\usepackage{amssymb}
\usepackage{mathrsfs}
\usepackage{xcolor,graphicx}
\usepackage[caption=false]{subfig}
\usepackage[toc,page]{appendix}
\usepackage[section]{placeins}
\usepackage{breqn}
\usepackage[T1]{fontenc}

\begin{document}

\title{On the fractal dimension of turbulent black holes}

\author{John Ryan Westernacher-Schneider}
\email{jwestern@uoguelph.ca}
\affiliation{Department of Physics \\
  University of Guelph \\
  Guelph\, Ontario N1G 2W1\, Canada}
\affiliation{Perimeter Institute for Theoretical Physics \\
  31 Caroline Street North \\
  Waterloo\, Ontario N2L 2Y5\, Canada}


\begin{abstract}
We present measurements of the fractal dimension of a turbulent asymptotically anti-deSitter black brane reconstructed from simulated boundary fluid data at the perfect fluid order using the fluid-gravity duality. We argue that the boundary fluid energy spectrum scaling as $E(k)\sim k^{-2}$ is a more natural setting for the fluid-gravity duality than the Kraichnan-Kolmogorov scaling of $E(k) \sim k^{-5/3}$, but we obtain fractal dimensions $D$ for spatial sections of the horizon $H\cap\Sigma$ in both cases: $D=2.584(1)$ and $D=2.645(4)$, respectively. These results are consistent with the upper bound of $D=3$, thereby resolving the tension with the recent claim in~\cite{Adams:2013vsa} that $D=3+1/3$. We offer a critical examination of the calculation which led to their result, and show that their proposed definition of fractal dimension performs poorly as a fractal dimension estimator on $1$-dimensional curves with known fractal dimension. Finally, we describe how to define and in principle calculate the fractal dimension of spatial sections of the horizon $H\cap \Sigma$ in a covariant manner, and we speculate on assigning a `bootstrapped' value of fractal dimension to the entire horizon $H$ when it is in a statistically quasi-steady turbulent state.
\end{abstract}

\maketitle

\section{Introduction}\label{sec:intro}
In a certain regime, the existence of turbulence in the gravitational field was recently demonstrated in numerical simulations of a perturbed black brane in asymptotically anti-deSitter (AAdS) spacetime in~\cite{Adams:2013vsa}. Such behavior was expected on the basis of the work of~\cite{Carrasco:2012nf} and the fluid-gravity duality, which gives an approximate dual description of the bulk geometry in terms of a conformal fluid living on the conformal boundary of the spacetime (see eg. \cite{Bhattacharyya:2008jc,Bhattacharyya:2008mz,VanRaamsdonk:2008fp}, or~\cite{Hubeny:2011hd} for a review and further references). This duality has opened the door to cross-pollination between the fields of gravity and fluid dynamics (eg.~\cite{Oz:2010wz, Eling:2010vr,Carrasco:2012nf,Green:2013zba,Eling:2013sna,Adams:2012pj, Adams:2013vsa,adams2014dynamical,Eling:2015mxa,WS2015,JRWS:2017}), and even resulting in insights relevant to gravitational wave astrophysics~\cite{Yang:2014tla}.

Interestingly, in~\cite{Adams:2013vsa} it was argued that a $(3+1)$-dimensional AAdS-black brane spacetime in a turbulent quasi-steady state has an event horizon with fractal dimension $D = 3 + 1/3$. Although the intersection of the horizon $H$ with a spacelike slice $\Sigma$ has dimension $2$, in a turbulent state one expects a bumpy horizon exhibiting approximate self-similarity over some range of scales, and therefore a fractal dimension $D$ in the range $2 \leq D \leq 3$. Since the result $D=3+1/3$ of~\cite{Adams:2013vsa} lies above this range, it is in tension with this basic expectation. Indeed, since $\Sigma$ is Riemannian and connected, it can be regarded as a metric space where its distance function is defined as the infimum of lengths of paths connecting any two points. Therefore, since $H\cap \Sigma$ is embedded in it, its fractal dimension cannot exceed the dimension of $\Sigma$~\cite{falconer1986geometry}.

In this work we begin in Sec.~\eqref{sec:background} with a review of the relevant calculation in~\cite{Adams:2013vsa}, then in Sec.~\eqref{sec:criticism} we provide a critical examination. We suggest that their calculation does not use their proposed definition of fractal dimension, and so the fact that their result exceeds the upper bound $D=3$ does not necessarily invalidate their definition. Nonetheless, in Sec.~\eqref{sec:testcases} we argue using well-understood test cases of statistically self-similar 1-dimensional curves embedded in the Euclidean plane that their proposed definition of fractal dimension is not reliable as a fractal dimension estimator. Next, in Sec.~\eqref{sec:results} we present an alternative numerical calculation of the fractal dimension of a turbulent black brane using simulated data of the dual turbulent fluid and the fluid-gravity duality at lowest (perfect fluid) order. We do so over the inverse-cascade range of a weakly-compressible conformal fluid with two sets of data corresponding Kraichnan-Kolmogorov scaling of the energy spectrum $E(k) \sim k^{-5/3}$ as well as the scaling $E(k)\sim k^{-2}$ which emerges as the direct-cascade becomes well-resolved (and in the absence of large-scale friction)~\cite{Scott:2007,JRWS:2017}. We obtain fractal dimensions of $D\approx 2.58$ and $D\approx 2.65$ for each case, respectively. Lastly, in Sec.~\eqref{sec:discussion} we describe what would be required to define, and in principle compute, the fractal dimension covariantly.
%
%
\section{Background}\label{sec:background}
In this section we briefly review the argument presented in~\cite{Adams:2013vsa}, specializing to the case of a $(3+1)$-dimensional bulk spacetime. Further details can be found in that work.

In~\cite{Adams:2013vsa} it is proposed that the fractal dimension of the horizon be defined via the scaling of the horizon area course-grained on a scale $\delta x$. Writing the course-grained area as a Riemann sum of the intrinsic area elements $A \approx \Sigma_i \sqrt{\gamma(x_i)} \Delta^2 x_i$, with $\Delta^2 x_i \approx (\delta x)^2$ and $\gamma(x_i)$ the intrinsic metric determinant evaluated at the point $x_i$, one extracts the purported fractal dimension $D$ from the scaling $A \sim (\delta x)^{2-D}$. One can immediately see that this definition has some of the expected behaviour: i) if the intrinsic metric determinant is constant over the surface, then the course-grained area does not depend on $\delta x$ so we must have $D=2$ (a smooth surface), ii) as the surface becomes rough ($D>2$), the area grows faster as $\delta x$ decreases (and indeed becomes infinite as $\delta x \rightarrow \infty$).

The calculation of the fractal dimension begins by considering the Raychaudhuri equation of the null generators of the horizon in vacuum,
\begin{eqnarray}
\kappa \mathcal{L}_n \sqrt{\gamma} + \frac{1}{2}\frac{1}{\sqrt{\gamma}} (\mathcal{L}_n \sqrt{\gamma})^2 - \mathcal{L}^2_n \sqrt{\gamma} = \sqrt{\gamma} \Sigma^i_j \Sigma^j_i,
\end{eqnarray}
where $n$ is the null normal, $\mathcal{L}_n$ is a derivative along $n$, $\kappa$ is the non-affinity of $n$ as per the expression $n_a \nabla^a n^b = \kappa n^b$, and $\Sigma^i_j$ is the shear. The regime of validity of the fluid-gravity duality is that of slowly-varying fields, so one expects the higher order derivative terms $ - \mathcal{L}^2_n \sqrt{\gamma}$ and $\frac{1}{2}\frac{1}{\sqrt{\gamma}} (\mathcal{L}_n \sqrt{\gamma})^2$ to be subleading. Dropping those terms and integrating over the spatial section of the horizon yields
\begin{eqnarray}
\frac{dA}{dt} = \int d^2x \frac{\sqrt{\gamma}}{\kappa} \Sigma^i_j \Sigma^j_i = \int_0^\infty dk \mathcal{A}(t,k), \label{eq:integratearea}
\end{eqnarray}
where $\mathcal{A}$ is the isotropic power spectrum of the `rescaled' shear $\theta^i_j \equiv \sqrt[4]{\gamma/\kappa^2} \Sigma^i_{\phantom{i}j}$, and the last equality follows from the Plancherel theorem. In~\cite{Adams:2013vsa} it was observed in full numerical simulations of a (3+1)-dimensional turbulent AAdS-black brane spacetime that $\mathcal{A} \sim k^2 E(k)$, at least in the regime of the simulations, where $E(k)$ is the isotropic velocity power spectrum of the boundary fluid. By assuming Kraichnan-Kolmogorov scaling $E(k) \sim k^{-5/3}$ over a the inverse-cascade range, one has $\mathcal{A} \sim k^{1/3}$ there. By inserting a large wavenumber cutoff $k_{\textrm{max}} \sim 1/\delta x$ in the wavenumber integral in Eq.~\eqref{eq:integratearea} one finds $dA/dt \sim k_{\textrm{max}}^{4/3} \sim (\delta x)^{-4/3}$ for $k_{\text{max}}$ in a sufficiently wide inverse-cascade range. Matching this to the scaling $(\delta x)^{2-D}$ finally yields $D=3+1/3$. 

One may object that the scaling $(\delta x)^{2-D}$ is to be applied to $A$, not $dA/dt$. However, in the quasi-steady state of a turbulent fluid forced at scale $k_f$ with inertial range scaling extending to a large scale $k_{\text{IR}}$, the spectrum $E(k)$ is well-approximated by piecewise power laws with the inertial range portion over $k \in (k_{IR},k_f)$ unchanging except that $k_{\text{IR}}$ decreases with time (see eg. \cite{fontane2013}). I.e. the inertial range becomes larger with time, but the spectrum over that range does not change. Plugging such a piece-wise power-law into the right-hand side of Eq.~\eqref{eq:integratearea} with UV cutoff $k_{\text{max}} \in (k_{IR},k_f)$ allows one to perform both the wavenumber and time integration explicitly. Thus the piecewise power-law model of $E(k)$ relevant to turbulent flows in quasi-steady state implies that $A$ and $dA/dt$ scale in the same way with the UV cutoff $k_{\text{max}}$, for $k_{\text{max}}$ sufficiently large. In the following section, we identify other possible sources of problems with the calculation.
%
%
\section{Critical examination}\label{sec:criticism}
We begin by noting what it means in position space to insert the small-scale cutoff $k_{\text{max}}$ in the wavenumber integral in Eq.~\eqref{eq:integratearea}. In order to do this we must write $\mathcal{A}(k,t)$ explicitly~\cite{Adams:2013vsa}:
\begin{eqnarray}
\mathcal{A}(t,k) \equiv \frac{\partial}{\partial k} \int_{|\boldsymbol{k}^\prime| \leq k} \frac{d^2k^\prime}{(2\pi)^2} \bar{\theta}^{*i}_{\phantom{*i}j} (t,\boldsymbol{k}^\prime) \bar{\theta}^j_{\phantom{j}i} (t,\boldsymbol{k}^\prime), \label{eq:curvaturespectrum}
\end{eqnarray}
where $\bar{\theta}^i_{\phantom{i}j} (t,\boldsymbol{k})$ is the Fourier transform of the rescaled horizon extrinsic curvature, $\bar{\theta}^i_{\phantom{i}j} (t,\boldsymbol{k}) = \int d^2x e^{-i\boldsymbol{k} \cdot \boldsymbol{x}} \theta^i_{\phantom{i}j} (t,\boldsymbol{x})$. Eq.~\eqref{eq:curvaturespectrum} can be rewritten as
\begin{eqnarray}
\mathcal{A}(t,k) &=& \frac{\partial}{\partial k} \int_0^k dk^\prime \int_0^{2\pi} d\phi \frac{k^\prime}{(2\pi)^2} \bar{\theta}^{*i}_{\phantom{*i}j} (t,\boldsymbol{k}^\prime) \bar{\theta}^j_{\phantom{j}i} (t,\boldsymbol{k}^\prime) \nonumber\\
&=& \left[ \int_0^{2\pi} d\phi \frac{k^\prime}{(2\pi)^2} \bar{\theta}^{*i}_{\phantom{*i}j} (t,\boldsymbol{k}^\prime) \bar{\theta}^j_{\phantom{j}i} (t,\boldsymbol{k}^\prime) \right]_{k^\prime = k} \nonumber\\
&=& k \int_0^{2\pi} \frac{d\phi}{(2\pi)^2} \bar{\theta}^{*i}_{\phantom{*i}j} (t,\boldsymbol{k}) \bar{\theta}^j_{\phantom{j}i} (t,\boldsymbol{k}).
\end{eqnarray}
Therefore the wavenumber integral in Eq.~\eqref{eq:integratearea} is just the integral of $\bar{\theta}^{*i}_{\phantom{*i}j} \bar{\theta}^{j}_{\phantom{j}i}/(2\pi)^2$ over all of Fourier space. Furthermore, note that integrating over $k\in (0,k_{\text{max}})$ is the same as multiplying by a step function kernel $\Theta(k_{\text{max}}-k)$ and then integrating over $k \in (0,\infty)$, and writing it in this way allows us to see the meaning of the cutoff in position space as follows:
\begin{eqnarray}
\int \frac{d^2k}{(2\pi)^2} \bar{\theta}^{*i}_{\phantom{*i}j} (t,\boldsymbol{k}) \bar{\theta}^j_{\phantom{j}i} (t,\boldsymbol{k}) \Theta(k_{\text{max}}-k) &=& \int \frac{d^2k}{(2\pi)^2} \left( \int d^2x\; e^{i\boldsymbol{k} \cdot \boldsymbol{x}} \theta^{i}_{\phantom{i}j} (t,\boldsymbol{x}) \right) \left( \int d^2x^\prime e^{-i\boldsymbol{k} \cdot \boldsymbol{x}^\prime} \theta^j_{\phantom{j}i} (t,\boldsymbol{x}^\prime) \right) \Theta(k_{\text{max}}-k) \nonumber\\
&=& \int d^2x\; d^2x^\prime \left[ \int \frac{d^2k}{(2\pi)^2} \Theta(k_{\text{max}}-k) e^{-i\boldsymbol{k} \cdot (\boldsymbol{x} - \boldsymbol{x}^\prime)} \right] \theta^{i}_{\phantom{i}j}(t,\boldsymbol{x}) \theta^j_{\phantom{j}i}(t,\boldsymbol{x}^\prime) \nonumber\\
&=& \int d^2x\; d^2x^\prime k_{\text{max}} \frac{J_1(\pi k_{\text{max}} |\boldsymbol{x}-\boldsymbol{x}^\prime | )}{|\boldsymbol{x}-\boldsymbol{x}^\prime |} \theta^{i}_{\phantom{i}j}(t,\boldsymbol{x}) \theta^j_{\phantom{j}i}(t,\boldsymbol{x}^\prime) \nonumber\\
&\equiv & \int d^2x\; \theta^i_{\phantom{i}j} (t,\boldsymbol{x}) \left\langle \theta^j_{\phantom{j}i} (t,\boldsymbol{x}) \right\rangle_{\delta x}, \label{eq:meaningpositionspace}
\end{eqnarray}
where $J_1$ is the Bessel function of the first kind, and we have defined $\left\langle \cdot \right\rangle_{\delta x}$ as a spatial coarse-graining operation at scale $\delta x \sim 1/k_{\text{max}}$ (in this case with an isotropic kernel $k_{\text{max}} J_1(\pi k_{\text{max}} |\boldsymbol{x}-\boldsymbol{x}^\prime | ) / |\boldsymbol{x}-\boldsymbol{x}^\prime |$).

We thus arrive at our first concern: the relationship between Eq.~\eqref{eq:meaningpositionspace} and the proposed coarse-graining $A \approx \Sigma_i \sqrt{\gamma(x_i)} (\delta x)^2$ is unclear. The latter is a Riemann sum, which if applied to the Raychauduri Eq.~\eqref{eq:integratearea} would yield $\Sigma_i \theta^j_{\phantom{j}l}(x_i) \theta^l_{\phantom{l}j}(x_i) (\delta x)^2 $. The summand could be viewed as a coarse-graining of both factors of the rescaled horizon extrinsic curvature $\theta$, whereas in Eq.~\eqref{eq:meaningpositionspace} only one factor of $\theta$ is coarse-grained. Thus, even if the scaling $A \approx \Sigma_i \sqrt{\gamma(x_i)} (\delta x)^2 \sim (\delta x)^{2-D}$ correctly captures the fractal dimension $D$, it is unclear whether the calculation performed in~\cite{Adams:2013vsa} uses it.

%
%
\subsection{Comparing methods on $1$-dimensional test cases}\label{sec:testcases}

Next, we argue that the Riemann sum approach $A \approx \Sigma_i \sqrt{\gamma(x_i)} (\delta x)^2 \sim (\delta x)^{2-D}$ is a poor fractal dimension estimator. We consider the 1-dimensional version of this, $L \approx \Sigma_i \sqrt{\gamma(x_i)} \delta x \sim (\delta x)^{1-D}$, applied to three different noise curves in the Euclidean plane whose fractal dimensions are known. We refer to this proposed method of determining the fractal dimension of a curve as the `intrinsic metric method'. Despite a strong resemblance, the intrinsic metric method of approximating the length of the curve is distinct from the `compass' or `ruler' method appearing in the pioneering study of coastline lengths~\cite{mandelbrot1967long}, since the former involves approximating the curve by its tangents at the points $x_i$, which are line segments of unequal length and whose end points do not necessarily lie on the curve. Each curve is defined by a function $f(x)$, and therefore has an intrinsic metric induced by the Euclidean metric of its embedding space whose determinant is $1+(\partial_x f)^2$. Thus we can compute coarse-grained versions of the length of the curve as $L \approx L_{\delta x} \equiv \Sigma_i \sqrt{1+(\partial_x f)^2}|_{x=x_i} \delta x$ and then compare with the expected scaling $(\delta x)^{1-D}$. For comparison, we estimate the fractal dimension of the same curves using the madogram method described in~\cite{gneiting2012estimators}. The madogram is defined as $\gamma_1(r) = (1/2) \left\langle \vert f(x) - f(x+r)\vert \right\rangle$, where $\left\langle \cdot \right\rangle$ denotes a spatial average. The madogram is expected to scale as $r^{2-D}$.

\begin{figure}[h!]
\centering
\hbox{\hspace{0cm}\includegraphics[width=\textwidth]{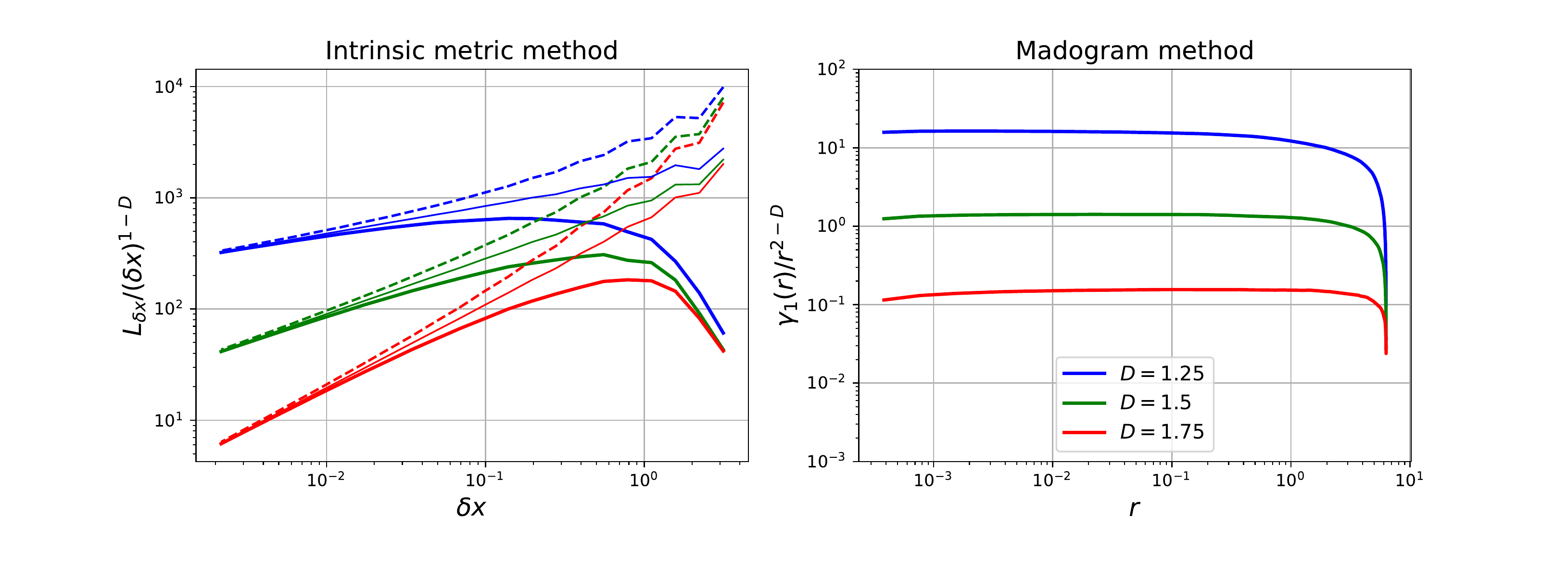}}
\caption{A comparison between the intrinsic metric and madogram methods for estimating the fractal dimension of $1$-dimensional noise with different fractal dimensions $D=1.25$ (blue), $D=1.5$ (green), $D=1.75$ (red). (Left): The coarse-grained length of each noise curve as a function of the coarse-graining scale $\delta x$, using the intrinsic metric method. Each plot is compensated by the expected scaling $(\delta x)^{1-D}$. The thick solid line corresponds to taking the minimum length over $10^3$ shifts of the sampling positions, while the thin solid line corresponds to taking the maximum, and the dashed line corresponds to taking the median. There is no discernible range of $\delta x$ over which the expected scaling is observed, so we conclude that this is method is not an accurate fractal dimension estimator. (Right): By contrast, the madogram $\gamma_1 (r)$ plotted as a function of $r$, compensated by the expected scaling $r^{2-D}$, for the same three noise curves. The expected scaling is clearly evident over a wide range of $r$.} \label{fig:noise_test}
\end{figure}

\begin{figure}[h!]
\centering
\hbox{\hspace{0.9cm}\includegraphics[width=0.9\textwidth]{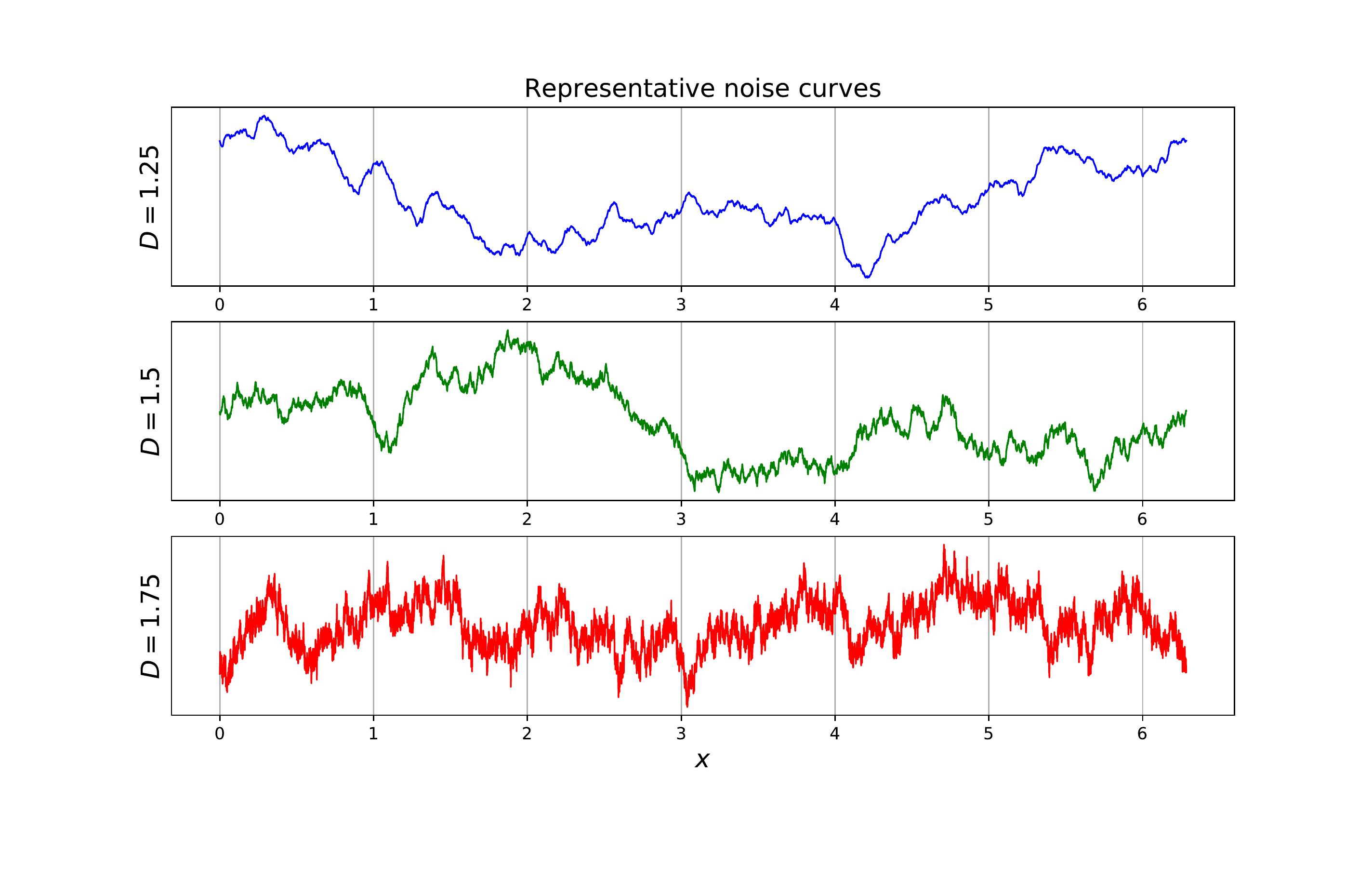}}
\caption{Representative noise curves with fractal dimensions $D=1.25$ (Top), $D=1.5$ (Middle), $D=1.75$ (Bottom). The $D=1.5$ case corresponds to Brownian noise. These curves have power spectra scaling as $k^{-\beta}$ for $\beta = 2.5$, $\beta = 2$, and $\beta = 1.5$, respectively.} \label{fig:noise_curves}
\end{figure}

Fig.~\eqref{fig:noise_test} shows a comparison between the intrinsic metric and madogram methods for estimating the fractal dimension of three noise curves with $D=1.25$ (blue), $D=1.5$ (green), $D=1.75$ (red). Such noise curves have power spectra scaling as $k^{-\beta}$ for $\beta = 2.5$, $2$, $1.5$, respectively. For the range $\beta \in [1,3]$ a topologically $d$-dimensional surface has a fractal dimension $D$ related to the spectral exponent by the approximate relation $D=(2d+3-\beta)/2$~\cite{voss1986}. For $\beta \leq 3$ the surface is sufficiently smooth that the fractal dimension equals its topological dimension, $D=d$, whereas for $\beta \leq 1$ it saturates to $D=d+1$.  In Fig.~\eqref{fig:noise_curves} we display representative curves with fractal dimensions $D=1.25$ (blue, Top), $D=1.5$ (green, Middle), and $D=1.75$ (red, Bottom). 

An ensemble of size $N=100$ is generated for each noise curve, and the estimators $L_{\delta x}$ and $\gamma_1 (r)$ are computed for each member and then averaged over the ensemble. In the intrinsic metric method, it is insufficient to attempt a single set of sampling locations for a given $\delta x$. Instead, we start with the zeroth set of sampling locations $\{ x_i \} = \{0, \delta x, 2\delta x, ... \}$, but also try $999$ additional sets related to the first by a translation $(m/1000)\delta x$ for the $m$th set, for a total of $1000$ sets. This results in $1000$ length estimates $L_{\delta x}$ for each $\delta x$, and we consider taking the minimum, maximum, or median length estimates, shown in Fig.~\eqref{fig:noise_test} (Left) in thick solid, thin solid, and dashed curves, respectively. Such ``shifts'' are an essential part of many fractal dimension estimating algorithms. Box-counting, for example, requires finding the \emph{minimum} number of boxes that cover the object, so many shifts of the box grid must be tried in order to obtain an accurate estimate. \emph{A priori} we do not know whether to take the minimum, maximum, or median estimate of the length $L_{\delta x}$. Different methods for estimating the fractal dimension have different conventions, for example the `compass' or `ruler' dimension~\cite{mandelbrot1967long} takes the maximum length, `box-counting' takes the minimum number of boxes~\cite{barnsley2014fractals}, and 'line transect variogram' methods applied to a surface take the median result from the transects~\cite{gneiting2012estimators}. However, as Fig.~\eqref{fig:noise_test} (Left) shows, none of the three possibilities yield the expected scaling $(\delta x)^{1-D}$ over any discernable range of $\delta x$. We note that taking the average or the median yields nearly identical curves (thin solid).

%
%
\section{Results}\label{sec:results}
Using the numerical code described in~\cite{JRWS:2017}, we evolve a $(2+1)$-dimensional conformal perfect fluid with equation of state $P=\rho/2$ on a $2\pi$-periodic domain with $2048^2$ points. The energy momentum tensor of the fluid is $T_{ab} = (3/2)\rho u_a u_b + (1/2)\rho \eta_{ab}$, with $u^a = \gamma (1, \boldsymbol{v})$ and $\gamma$ the Lorentz factor. The fluid is evolved from rest $\rho=1$, $\boldsymbol{v}=0$, and turbulence is induced and sustained by a random external force with homogeneous, isotropic, Gaussian white-noise-in-time statistics. The external force has support in a narrow band of wavenumbers around $k_f$. Further details can be found in~\cite{JRWS:2017}.

\begin{figure}[h!]
\centering
\hbox{\hspace{0.7cm}\includegraphics[width=0.9\textwidth]{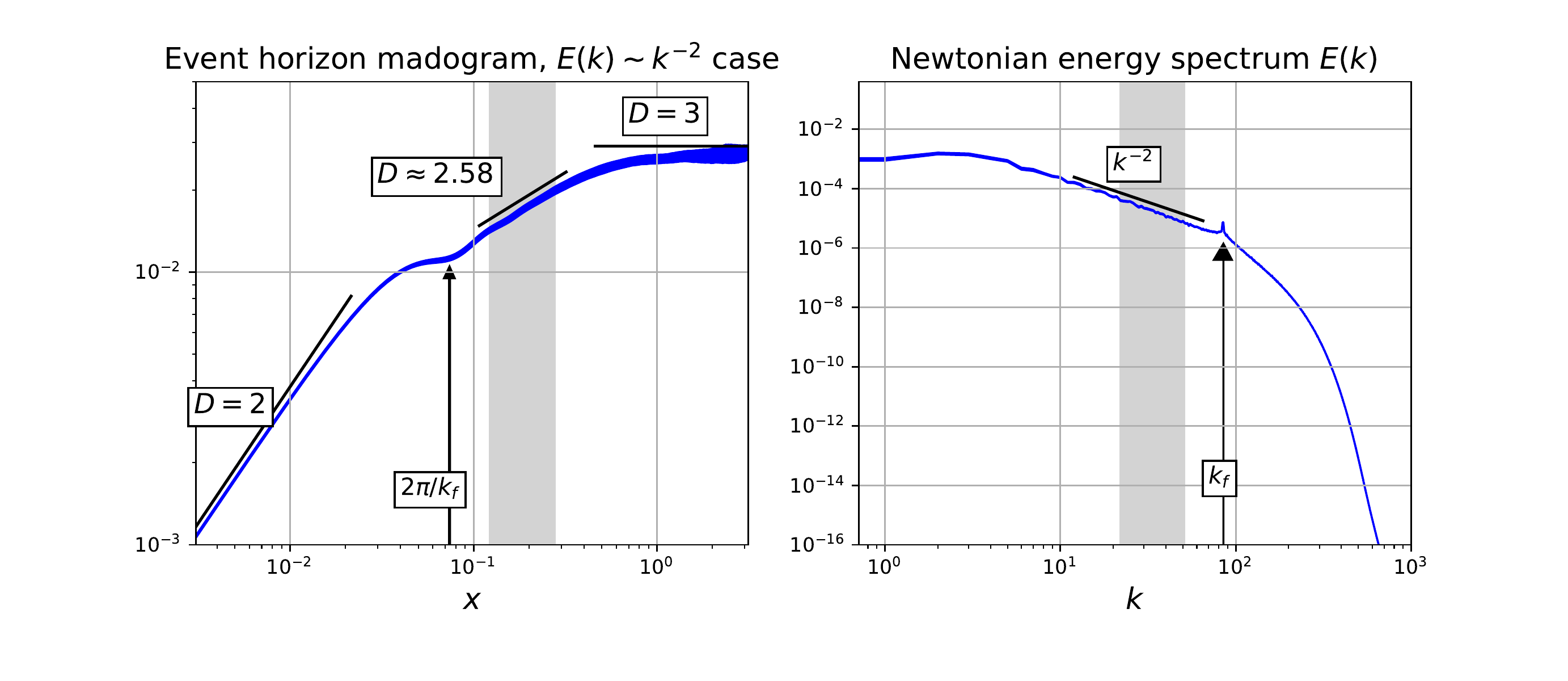}}
\caption{Event horizon madogram (Left) and corresponding boundary fluid isotropic Newtonian kinetic energy spectrum (Right) for the case with $k_f = 85$. The thickness of each plot corresponds to the $\sqrt{N}$ statistical uncertainty. (Left): The madogram yields a fractal dimension of $2$ at small scales, thus agreeing with the topological dimension. This is expected since the horizon is not a true fractal, i.e. it does not exhibit rough structure down to arbitrarily small scales. Above the forcing scale $2\pi/k_f$, a scaling range is observed with $2.584(1)$, where we have indicated the statistical uncertainty in brackets $()$. The range of $x$ over which we fit a power-law is indicated as the shaded grey region, and the corresponding range of wavenumbers is also indicated (Right). At the largest scales, the madogram saturates to $D=3$, which corresponds to the flow resembling white noise there (i.e. $E(k) \sim$ constant). (Right): The isotropic Newtonian specific kinetic energy spectrum $E(k) = \pi \left\langle |\hat{v}|^2 \right\rangle (k)$. A power-law of $k^{-2}$ is shown for reference, and the forcing scale $k_f$ is indicated with an arrow.} \label{fig:scott_madogram}
\end{figure}

\begin{figure}[h!]
\centering
\hbox{\hspace{0.7cm}\includegraphics[width=0.9\textwidth]{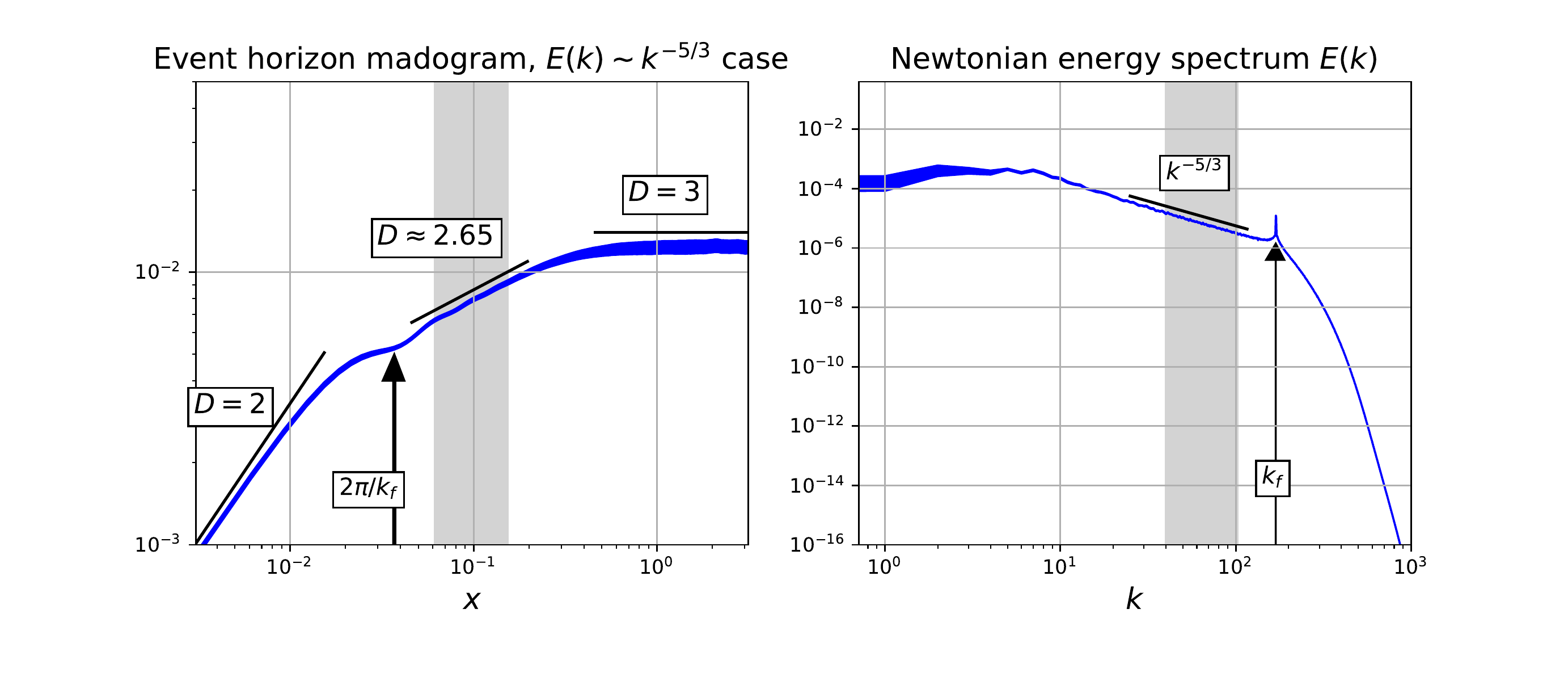}}
\caption{The corresponding plots as in Fig.~\eqref{fig:scott_madogram}, but for the boundary fluid exhibiting Kraichnan-Kolmogorov scaling of the energy spectrum, $E(k) \sim k^{-5/3}$. In this case, the measured fractal dimension is $D=2.645(4)$ over the inverse-cascade range. This value is slightly higher than the case with $E(k) \sim k^{-2}$, which is expected since the flatter spectrum indicates rougher structure. }\label{fig:kraichnan_madogram}
\end{figure}

An inverse-cascade range develops, and since we do not implement any large-scale energy sinks, the resulting flow is referred to as being in a quasi-steady state~\cite{Scott:2007}. For two separate cases with $k_f = 85$ and $k_f = 170$, we generate an ensemble of $20$ flows and perform analysis on snapshots prior to the energy piling up at the scale of the box. As displayed in Fig.~\eqref{fig:scott_madogram} (Right), the $k_f=85$ case yields an isotropic Newtonian specific kinetic energy spectrum $E(k) \sim k^{-2}$, as found in~\cite{Scott:2007} in the incompressible case and confirmed in~\cite{JRWS:2017} for a conformal fluid in the weakly-compressible regime.\footnote{It was found in~\cite{Scott:2007} that the spectrum steepens to $\sim k^{-2}$ when $k_{\text{max}}/k_f \gtrsim 16$, where $k_{\text{max}} \equiv N/3$ and $N$ is the number of points on the grid. In our simulations this would correspond to $k_f \approx 41$, but in their case regular 2nd-order viscosity was used, whereas we use 4th-order dissipation. Thus, we are able to achieve the $k^{-2}$ spectrum with a much larger $k_f$ because our dissipation operates at larger wavenumbers.} The $k^{-2}$ scaling is associated with both a well-resolved direct cascade and an absence of large-scale friction. Since the regime of validity of the fluid-gravity duality is that of an arbitrarily high Reynolds number and no large-scale friction, we argue that this spectrum corresponds to the natural setting for the dual spacetime. However, for comparison we also consider the $k_f = 170$ case, where the force is active deeper into the dissipation range, and which yields the traditional Kraichnan-Kolmogorov scaling $E(k) \sim k^{-5/3}$, as displayed in Fig.~\eqref{fig:kraichnan_madogram}.

We applied the madogram method to $x$- and $y$-transects of the event horizon. The fractal dimension $D$ of the horizon is then obtained by extracting $D_{\text{transect}} \in [1,2]$ from the median madogram of each topologically $1$-dimensional transect, and then writing $D = D_{\text{transect}} +1$. Such a prescription is valid for surfaces exhibiting statistical self-similarity, and its performance was evaluated extensively in~\cite{gneiting2012estimators}. In Figs.~\eqref{fig:scott_madogram} and~\eqref{fig:kraichnan_madogram} (Left) we display the median madogram over all transects of the horizon (herein referred to as the `event horizon madogram'), as applied to the radial coordinate position of the event horizon, $r_{+}(x^c) = 4\pi T(x^c)/3$, for the perturbed boosted AAds-black brane metric at perfect fluid order,
\begin{eqnarray}
ds^2 = -2 u_a(x^c) dx^a dr - \frac{r^2}{R^2}(1-\frac{r_{+}^3(x^c)}{r^3})u_a(x^c) u_b(x^c) dx^a dx^b + \frac{r^2}{R^2}(\eta_{ab} + u_a(x^c) u_b(x^c)) dx^a dx^b,
\end{eqnarray} 
where the indices $(a,b)$ run over the `boundary' directions $(t,x,y)$ only, $R$ is the AdS length scale (which we set to 1), $u_a$ is the boost $4$-velocity, and $\eta_{ab}$ is the $(2+1)$-dimensional Minkowski metric. For the $(2+1)$-dimensional boundary conformal fluid, $T = \rho^{1/3}$. The perturbations are imagined to be slowly-varying with respect to the boundary directions, which will solve Einstein's equations with arbitrary accuracy in the perfect fluid limit if $u^a$ and $T$ evolve according to conformal hydrodynamics on the boundary. Error estimates have been obtained for solutions constructed from particular boundary fluid data in~\cite{Adams:2013vsa} via direct comparison with full GR simulations, showing agreement at the $1\%$ level (see also~\cite{adams2014dynamical} for error estimates which do not use full GR simulations). 

Fig.~\eqref{fig:scott_madogram} shows the case with $E(k) \sim k^{-2}$ and Fig.~\eqref{fig:kraichnan_madogram} shows the case with $E(k) \sim k^{-5/3}$. The thickness of each plot indicates the $\sqrt{N}$ statistical uncertainty associated with the ensembles. At small scales $x \ll 2\pi/k_f$, the horizons have fractal dimension $2$, which agrees with their topological dimension. This is expected since rough structure does not persist down to arbitrarily small scales. For a range of scales greater than the forcing scale $2\pi/k_f$, power-law behavior is observed in both cases. A least-squares power-law fit over the grey shaded intervals yield a fractal dimension of $D=2.584(1)$ and $D=2.645(4)$ for the cases $E(k) \sim k^{-2}$ and $E(k) \sim k^{-5/3}$, respectively, with $\sqrt{N}$ uncertainties indicated. The corresponding fitting interval in Fourier space is indicated on the plots of the energy spectra (Right). The madograms saturate at $D=3$ at large scales, beyond the inertial range scale, which is due to the flow resembling white noise at those scales (i.e. $E(k) \sim$ constant).

%
%
\section{Covariant construction of fractal dimension}\label{sec:discussion}

Many methods exist for calculating the fractal dimension of a set $F$ embedded in an ambient metric space $(\mathcal{M},d)$, where $d$ is a distance function $d: \mathcal{M}\times\mathcal{M} \rightarrow \mathbb{R}$ which is symmetric $d(x,y)=d(y,x)$ and satisfies $d(x,y) = 0 \Leftrightarrow x=y$ and the triangle inequality $d(x,y)+d(y,z) \geq d(x,z)$.  See eg.~\cite{gneiting2012estimators} for a comparison of many fractal dimension estimator algorithms when the metric space is Euclidean, or~\cite{falconer2004fractal,barnsley2014fractals} for strictly mathematically equivalent definitions. For illustrative purposes, we will focus on the box-counting method in this section, where one defines $N(\epsilon)$ to be the minimum number of boxes of size $\epsilon$ in the embedding space required to completely cover the set $F$, and then computes the fractal dimension as $\lim_{\epsilon \rightarrow 0} \log{(N(\epsilon))}/\log{(1/\epsilon)}$. The box-counting method is known to be diffeomorphism-invariant but not homeomorphism-invariant \cite{ott1984dimension} in the strict $\epsilon \rightarrow 0$ limit. The covering need not use boxes; indeed, any sets $U_i$ of diameter $\epsilon \equiv |U_i| = \sup \{ d(x,y): x,\: y\in U_i \} $ yield the same result~\cite{falconer1986geometry}.

In practical applications one does not take the $\epsilon \rightarrow 0$ limit, but instead fits a power-law to $N(\epsilon)\sim \epsilon^{-D}$ over some finite range $\Delta\epsilon = (\epsilon_{\text{UV}}, \epsilon_{\text{IR}})$. If the covering sets are not constructed covariantly, then any such fitting over $\Delta \epsilon$ would be subject to coordinate ambiguity, since the set in question could be made to appear smooth over the scale $\Delta \epsilon$ via a judicious choice of coordinates. In the example of a Brownian noise curve with $D=1.5$, described by $f(x)$ in Cartesian coordinates in Euclidean space, one could make the coordinate transformation ($\tilde{y} = \left\langle f(x) \right\rangle_{\Delta \epsilon}$, $\tilde{x} = x$), where $\left\langle f(x) \right\rangle_{\Delta \epsilon}$ is $f(x)$ with all modes outside the range of scales $\Delta \epsilon$ filtered out. Counting coordinate boxes over the scales $\Delta \epsilon$ would then give the incorrect result $D\approx 1$. Thus, it is important to construct the covering sets in a diffeomorphism-invariant way when performing a fit over the range of scales $\Delta \epsilon$.

When defining fractal dimension over a Riemannian manifold, one natural covariant choice of covering sets are geodesic balls $B(x,\epsilon/2)$, constructed by taking the union of all geodesics of length $\epsilon/2$ emanating from the point $x$. The event horizon $H\cap \Sigma$ on the slice $\Sigma$ is embedded isometrically both in $(\Sigma,h)$ and $(\mathcal{M},g)$, where $g$ is the full spacetime metric and $h$ is the induced metric on $\Sigma$. Since geodesic paths in $\Sigma$ need not be geodesic in $\mathcal{M}$, one is faced with the choice of whether to cover $H\cap\Sigma$ with geodesic balls in $\Sigma$ or in $\mathcal{M}$\footnote{Since $\mathcal{M}$ has a Lorentzian metric signature, a geodesic ball as defined above would contain the entire light cone of the central point $x$ since the length along null paths is zero. Thus, in this case we can instead define the geodesic ball $B_{\text{SL}}(x,\epsilon/2)$ as the union of only those spacelike geodesic paths of length $\epsilon/2$ emanating from $x$ which intersect $H\cap\Sigma$ at a point $y\neq x$.}. But note that the slice $\Sigma$ itself could be deformed at points off of $H\cap\Sigma$ to yield different geodesic paths while sharing the point set $H\cap \Sigma$. Thus, constructing the geodesic balls in $\Sigma$ would yield a fractal dimension which is not solely a property of $H\cap\Sigma$, but rather dependent on the arbitrary choice of slicing away from $H\cap\Sigma$. For this reason, we advocate using geodesic balls constructed in the full spacetime $\mathcal{M}$ (with a suitable redefinition - see footnote$^2$).

Computed in this way, a geodesic ball-counting procedure over a range of scales $\Delta \epsilon$ would yield a fully covariant estimate of the fractal dimension of any given spatial section of the event horizon. Furthermore, recall that our ensembles of event horizons considered in Sec.~\eqref{sec:results} yield roughly the same fractal dimension. It is often observed that cross-sections of $D$-dimensional fractals or statistically self-similar objects themselves have a fractal dimension of $D-1$ (see eg.~\cite{geologybook} for geological examples). Given our measurement of $D_{H\cap\Sigma}\approx 2.58$ in Sec.~\eqref{sec:results} for the $E(k)\sim k^{-2}$ case, this suggests that in a quasi-steady turbulent state the entire horizon $H$ can be assigned a fractal dimension of $D_H \approx 3.58$ (or $D_H \approx 3.65$ for the Kraichnan-Kolmogorov case $E(k)\sim k^{-5/3}$). However, the mathemetical meaning of this is not clear since $H$ is a null hypersurface embedded in a Lorentzian manifold $\mathcal{M}$, so there is difficulty in defining the diameter of covering sets in a covariant way.

Numerically implementing the procedure described in this section would be expensive, since one would have to integrate a large number of geodesics from a given point to construct a geodesic ball, do so for many geodesic balls to find a covering, and do this for many possible coverings to find the minimal one. In the current work we have not followed a covariant procedure like this. Many others have not either (eg.~\cite{cornish1996time,cornish1996chaos,cornish1997mixmaster,frolov1999chaotic,lehner2011final}), some opting instead to point out the diffeomorphism-invariance of box-counting in the $\epsilon \rightarrow 0$ limit while only fitting over a finite range $\Delta \epsilon$. It would be interesting to see how much these results change when done covariantly. 

Alternatively, it is plausible that the fractal dimension will not depend sensitively on the embedding space if, in a region around the surface, one has well-separated scales over which the surface and the embedding space vary. If this is true, once could obtain an approximate covariant result by embedding the surface isometrically in Euclidean space, and then applying a standard fractal dimension estimator. We attempted to embed the turbulent horizon isometrically in $\mathbb{E}^3$, without success. Indeed, the existence of such a (global) embedding is only guaranteed if the Gaussian curvature is positive over the entire surface, and may or may not exist otherwise. It has been observed~\cite{nollert1996visualization} that even a sufficiently rapidly-rotating Kerr black hole horizon does not have a global embedding in $\mathbb{E}^3$, since the Gaussian curvature becomes negative at the poles. We have computed the Gaussian curvature using our fluid data from Sec.~\eqref{sec:results}, and observed that it changes sign over the domain as rapidly as the external force. Thus, we believe it is highly unlikely that there exists a global embedding into $\mathbb{E}^3$ for arbitrary turbulent horizons in the regime of the fluid-gravity duality, although Euclidean embeddings are guaranteed to exist in sufficiently high dimensions.
%
%

\section{Conclusions}
In this work we provided a critical examination of the calculation in~\cite{Adams:2013vsa} which led to the claim that topologically $d$-dimensional turbulent AAdS-black brane horizons $H\cap\Sigma$ embedded in a $(d+1)$-dimensional Riemannian space $\Sigma$ have a fractal dimension $D=d+4/3$, exceeding the upper bound of $d+1$. We offered an alternative numerical computation of $D$ when $d=2$, and discussed issues surrounding the covariance of that quantity. 

In particular, we argued using well-understood test cases of $1$-dimensional noise curves that the proposed definition of fractal dimension in~\cite{Adams:2013vsa}, $A_{\delta x} = \sum_i \sqrt{\gamma(\boldsymbol{x}_i)} (\delta x)^2 \sim (\delta x)^{2-D}$, when specialized to topologically $1$-dimensional objects, performs poorly as a fractal dimension estimator. We emphasize that this is not a proof that the definition fails in the strict $\delta x \rightarrow 0$ limit for genuine fractals, but since the proposed application is on statistically self-similar surfaces which do not exhibit rough structure down to arbitrarily small scales, the performance of this proposal as a fractal dimension estimator is relevant. Furthermore, we argued that the calculation in~\cite{Adams:2013vsa} may not be using their proposed definition at all (so their result of $D=d+4/3$ alone does not necessarily invalidate their proposed definition, hence our separate evaluation of the definition on noise curves of known fractal dimension).

Using simulated turbulent conformal fluid flows in the quasi-steady state regime, we constructed snapshots of the turbulent event horizon using the fluid-gravity duality at perfect fluid order. By applying a line transect madogram method~\cite{gneiting2012estimators} to the event horizon surface $r_{+}(x,y) = 4\pi T(x,y)/3$ in boosted ingoing Finkelstein coordinates, we obtained a fractal dimension for spatial sections of the horizon $H\cap\Sigma$ of $D=2.584(1)$ and $D=2.645(4)$ for the cases with the boundary spectrum $E(k)\sim k^{-2}$ and $E(k)\sim k^{-5/3}$, respectively. We argued that the former scaling, $E(k) \sim k^{-2}$, is a more natural setting for the fluid-gravity duality since it corresponds to the regime of infinite Reynolds number without large-scale dissipation of energy~\cite{Scott:2007,JRWS:2017}.

We also speculated that in the quasi-steady state regime, since the fractal dimension will statistically not depend on the particular time at which a spatial section of the horizon is considered, that the entire horizon $H$ could be assigned a `bootstrapped' fractal dimension of $D_H = D_{H\cap\Sigma} + 1$, although the strict mathematical meaning of this is not clear. Furthermore, we have not shown that $D_{H\cap\Sigma}$ is invariant with respect to deformations of the spatial section of the horizon, since we have only considered constant time slices in the ingoing Finkelstein coordinate.

%
%
\acknowledgements
We thank Neil Cornish, Zachary Vernon, Robie Hennigar, and Eric Poisson for helpful discussions. J.R.W.S. acknowledges support from OGS. 
Research at Perimeter Institute is supported through Industry Canada and by the Province of Ontario through the Ministry of Research \& Innovation.

\bibliography{fluidbib}

\end{document}